\documentclass[a4paper,12pt]{article}

\usepackage[mathscr]{eucal}
\usepackage{amsmath,amsfonts,amssymb,amsthm}
\usepackage{times}
\usepackage{hyperref}

\advance\textheight\topskip
\numberwithin{equation}{section} \makeatletter
\@addtoreset{equation}{section}

\renewcommand{\tilde}{\widetilde}
\renewcommand{\hat}{\widehat}

\newcommand{\bref}[1]{\textbf{\ref{#1}}}

\renewcommand{\d}{\partial}

\renewcommand{\leq}{\,{\leqslant}\,}

\newcommand{\inner}[2]{\langle #1{,}\,#2\rangle}
\newcommand{\bfinner}[2]{\bm{\langle} #1{,}\,#2\bm{\rangle}}
\newcommand{\binner}[2]{%
  {\langle}\kern-4.15pt{\langle}#1{,}\,#2{\rangle}\kern-4.15pt{\rangle}}

\newcommand{\ffrac}[2]{\raisebox{.5pt}%
  {\footnotesize$\displaystyle\frac{#1}{#2}$}\kern1pt}

\newcommand{\derham}{\boldsymbol{d}}

\newcommand{\dl}[1]{\frac{\partial}{\partial #1}}

\newcommand{\fR}{\mathbb{R}}

\def\cA{\mathcal{A}}

\def\cP{\mathcal{P}}

\def\div{{\rm Div}}

\def\BG-Poincare{Barnich:2009jy}
\def\Fedosov-book{Fedosov:1996fu}

\usepackage{youngtab}
\usepackage{ytableau}
\usepackage[numbers,sort&compress]{natbib}
 \usepackage[nottoc]{tocbibind}

\makeatletter
\g@addto@macro\bfseries{\boldmath}
\makeatother

\usepackage{multirow}

\usepackage{bm}

\usepackage{hhline}

\usepackage{physics}

\usepackage{tikz-cd}

\newcommand*\dif{\mathop{}\!\mathrm{d}}

\usepackage{mathtools}

\newcommand{\divw}{\operatorname{D}}

\newcommand{\youngw}{\operatorname{Y}}
\newcommand{\divwfullasym}{\mathfrak D}
\renewcommand{\div}{\operatorname{A}}

\newcommand{\divwcenter}{\operatorname{B}}

\usepackage{authblk}
\title{Conformal Lagrangians from the (formal) near boundary analysis of AdS gauge fields\vspace{0.5em}}
\author[a,b]{Alexander Chekmenev}
\author[a,c]{Maxim Grigoriev\vspace{0.75em}}
\affil[a]{Lebedev Institute of Physics,\protect\\
  Leninsky ave. 53, 119991 Moscow, Russia \vspace{1em}}
\affil[b]{Moscow Institute of Physics and Technology,\protect\\
  Institutskiy per. 7, Dolgoprudny, 141700 Moscow region, Russia \vspace{1em}}
\affil[c]{Institute for Theoretical and Mathematical Physics,\protect\\
  Lomonosov Moscow State University, 119991 Moscow, Russia}
\date{}                     
\begin{document}

\maketitle

\begin{abstract}
A simple generating procedure for Lagrangians of conformal gauge fields of mixed-symmetry type is presented. The construction originates from the analysis of the near-boundary behaviour of the associated AdS gauge fields
using the ambient space approach to leading boundary values. Manifestly ambient form of the Lagrangian is also obtained. 
As an illustration we apply the procedure to the simplest mixed-symmetry conformal gauge field, described by the two-row Young diagram, and derive the explicit component form of the respective Lagrangian.
\end{abstract}

\newpage

\tableofcontents

\section{Introduction}

Conformal higher spin gauge theories attract considerable attention because they give tractable examples of interacting Lagrangian theories that extend conformal gravity and are holographically related to higher spin gauge theories in AdS space of one dimension higher. The simplest example of conformal higher spin fields are totally symmetric gauge fields in even dimensional Minskowski space, which are known as Fradkin-Tseytlin fields. They provide a field content of the conformal higher spin gravity~\cite{Segal:2002gd,Tseytlin:2002gz}.

Conformal fields in $d$ dimensions are intimately related to their associated AdS fields living in $(d+1)$-dimensional AdS. More precisely, conformal fields can be identified as boundary values of the respective AdS fields.
Furthermore,  the Lagrangian for conformal fields can be derived as a logarithmically-divergent part of the effective action for the respective AdS fields~\cite{Aharony:1999ti, Henningson:1998gx,Skenderis:2002wp}. In the extension~\cite{Metsaev:2009ym} of this approach to higher spin theories one starts with the Lagrangian of higher spin gauge fields in the bulk.

It turns out that equations of motion of conformal fields can be inferred from the bulk dynamics without resorting to the Lagrangian formulation in AdS. More precisely, the conformally invariant equations on boundary values arise as conditions ensuring that the unconstrained boundary value can be lifted to an on-shell bulk field.  This interpretation dates back to the celebrated Fefferman-Graham~\cite{FG} construction. The extension of this approach to HS theories was proposed in~\cite{Bekaert:2012vt,Bekaert:2013zya,Bekaert:2017bpy} (see also~\cite{Grigoriev:2018mkp}) and resulted in a rather concise formulation of Fradkin-Tseytlin fields and their higher-depth generalizations. As far as conformal (gauge) fields of general symmetry type are concerned the generalization of the approach becomes somewhat inevitable because it leads to a concise and handful formulation of mixed symmetry conformal gauge fields~\cite{Chekmenev:2015kzf} at the level of equations of motion.

In contrast to the equations of motion, Lagrangians of generic conformal mixed-symmetry (gauge) fields can not be obtained from the Lagrangians of the respective AdS fields  simply because the latter are not yet known in the general case. However, in particular cases, where Lagrangian description in the bulk is available,  the standard strategy works~\cite{Alkalaev:2012ic}. Lagrangians for a rather general conformal (gauge) fields have been proposed in~\cite{Vasiliev:2009ck} from a different perspective and their interpretation in terms of the bulk dynamics remains somewhat unclear.

In this work we derive Lagrangian description of a wide class of conformal mixed-symmetry fields directly from their bulk dynamics. Although the Lagrangian description in the bulk is not available the bulk equations of motion (more precisely, their ambient space version) naturally provide us with the gauge-invariant kinetic operator defined on the leading boundary values of the bulk fields. Moreover we succeeded to build an inner product with respect to which the kinetic operator is formally self-adjoint and hence immediately gives us a local and gauge invariant Lagrangian. 
We show that at least in the simplest cases our Larangian is equivalent to the one of~\cite{Vasiliev:2009ck}. The advantage of our approach is that it makes the relation to bulk dynamics manifest and is conformally invariant by construction. Moreover, our Lagrangian admits a very simple reformulation in the ambient terms.

Additional motivation of this work has to do with already mentioned long-standing problem of Lagrangian formulation for general mixed-symmetry gauge fields in AdS. It is tempting to expect that this may help to lift the construction to the bulk, leading to the Lagrangian description of AdS fields.

The paper is organized as follows: In Section \bref{sec:ambient} we review the ambient description of AdS fields and fix a class of fields we work with.
In Section \bref{sec:boundary_values} we recall how this description leads to a concise formulation of the conformal equations of motion satisfied by the leading boundary values. In Section~\bref{sec:lagrangians} we prose the inner product that makes kinetic operator formally self-adjoint, giving a  gauge invariant Lagrangian. The ambient form of the Lagrangian is constructed in Section~\bref{sec:ambient-form}.   Appendix \bref{app:w_lift} contains some technical details of the construction.

\section{Ambient description of AdS fields} \label{sec:ambient}

Our approach to conformal Lagrangians originates from the ambient description of AdS gauge fields proposed in~\cite{Barnich:2006pc,Alkalaev:2009vm,Alkalaev:2011zv}. Here we mainly follow~\cite{Alkalaev:2011zv}.  This approach
is based on the extensive use of the ambient space. 

More specifically, tensor fields on $AdS_{d+1}$ are described in terms of tensor fields defined on the ambient space $\fR^{d+2}/\{0\}$, which is a pseudo-Eucledean space of signature $d+2$ with the origin excluded.  We use Cartesian cooridinates 
$X^A, \, A = 0, \ldots, d+1$ on the ambient space, where components of the metric are $\eta_{AB}=\eta(\frac{\d}{\d X^A},\frac{\d}{\d X^B})$. $AdS_{d+1}$ can be understood as hyperboloid $X^2=-1$ embedded in the ambient space.  The restriction of $o(d,2)$ transformations to the hyperboloid gives the algebra of inifinitesimal AdS isometries.
Although in this way one can not describe the most general
negative constant curvature spaces it turns out that the resulting description is covariant and hence applicable in the general case. This is because the ambient space construction is eventually implemented in the fiber of a suitable fiber bundle rather than in the space-time. This is achieved through the appropriate version of the parent formulation~\cite{Barnich:2006pc,Alkalaev:2009vm}.

To work with tensor fields on the ambient space we employ the language of generating functions. To this end we consider the algebra of polynomials in the auxiliary coordinates 
$P_i^A,\, i = 1, \ldots, n-1, A = 0, \ldots, d+1$.  This algebra contains ambient tensors with $n-1$ groups of totally symmetric indexes and hence is wide enough to contain all the relevant representations of $o(d-1,2)$.  We then introduce the ambient space function $\Phi$ with values in the algebra which we treat as a generating function for AdS fields. More precisely, the coefficients entering in the expansion of $\Phi(X,P_i)$ in $P^A_i$ are precisely tensor fields on the ambient space.

The ambient space tensor fields form a natural representation space of the algebra $o(d,2)$ of isometries. In terms of generating functions the $o(d,2)$-generators are represented as differential operators:
\begin{equation} \label{od2_algebra}
	J_{A B} = X_A \frac{\d}{\d X^B} + P_i^A \frac{\d}{\d P_i^B} - (A \leftrightarrow B).
\end{equation}
The space of ambient tensors is also a module over $sp(2n)$, which together with $o(d-1,2)$ module structure gives a standard setting of Howe duality. More precisely, respective groups centralise each other in this bimodule. 

In what follows we only need to introduce notation for the following $sp(2n)$ generators:
\begin{equation}
\begin{gathered}
	T^{i j} = \frac{\d}{\d P_i{}_A} \frac{\d}{\d P_j^A},\quad
  N_i{}^j = P_i^A \frac{\d}{\d P_j^A},\quad
  N_i = N_i{}^i,\quad
  N_X = X^A \frac{\d}{\d X^A},\\
  \Box = \frac{\d}{\d X_A} \frac{\d}{\d X^A},\quad
  S^i = \frac{\d}{\d P_i^A} \frac{\d}{\d X_A},\quad
  S^\dag_i = P_i^A \frac{\d}{\d X^A},\quad
  \bar S^\dag{}^i = X^A \frac{\d}{\d P_i^A}.
\end{gathered}
\end{equation}
They form a subalgebra of $sp(2n)$.

A generic mixed symmetry field of spin $\{s_1, s_2, \ldots, s_{n-1}\}$ (it is assumed that $s_1 \ge s_2 \ge \ldots \ge s_{n-1}$ and $n-1 \le \left[ \frac{d}{2} \right]$) on $(d+1)$-dimensional AdS space can be described by the following constraints, which depend on extra real parameter $\Delta$ and positive integer parameter $p\leq n-1$:

\paragraph{Purely algebraic constraints} These are generalized tracelessness, Young-symmetry and spin-weight conditions:
\begin{equation} \label{algebraic-constraints}
T^{ij} \Phi = 0, \qquad N_i{}^j \Phi = 0, \; i<j, \qquad N_i \Phi = s_i \Phi.
\end{equation}

\paragraph{Tangent constraints}
\begin{equation}
{\bar S^\dagger}{}^{\hat \alpha}\Phi=0\,,\quad \hat \alpha=p+1,\ldots,n-1
\end{equation}
their role is to reduce tensor in $d+2$ dimensions to (a collections of) tensors in $d+1$. 

\paragraph{Radial weight constraint}
\begin{equation}
(N_X + \Delta) \Phi = 0,
\end{equation}
where $\Delta$ is a parameter of the theory. Roughly speaking, this constraint fixes the radial dependence of $\Phi$.


\paragraph{Equations of motion (and partial gauges)} 
\begin{equation} \label{eom-and-pg}
\Box \Phi = 0, \qquad S^i \Phi = 0.
\end{equation}
In contrast to the above ones these are essentially differential constraints because they do involve $X^A$ derivatives along the hyperboloid and, being rewritten in terms of tensor fields on the hyperboloid, are precisely the equations of motion together with partial gauge conditions.

\paragraph{Gauge invariance} 
The above constraints in general describe a reducible system. Indeed, the space of fields satisfying the constraints has an invariant submodulde, which gives rise to the following linear gauge transformation:
\begin{equation}
\delta_\chi \Phi = S^\dag_\alpha \chi^\alpha, \qquad \alpha = 1, \ldots, p,
\end{equation}
where gauge parameters $\chi^\alpha$ satisfy the same constraints as $\Phi$ except those involving $N_X, N_i, N_i{}^j$ which are replaced by
\begin{equation}
(N_X + \Delta - 1) \chi^\alpha = 0,
\end{equation}
\begin{equation}
N_i \chi^\alpha = s_i \chi^\alpha - \delta^\alpha_i \chi^\alpha,
\end{equation}
\begin{equation} \label{ambient_Young}
N_i{}^j \chi^\alpha = - \delta^\alpha_i \delta^j_\beta \chi^\beta \quad i < j.
\end{equation}

\paragraph{Extra tangent constraint.} For $\Delta$ generic the above system is irreducible. But for special $\Delta$, namely such that $\Delta = t + p - s_p$, where $t \in \left\{ 1, 2, s_p - s_{p+1} \right\}$ the extra condition needs to be imposed for the system to be irreducible:
\begin{equation}
{\bar S^\dagger}{}^{t}\Phi=0.
\end{equation}

\subsection{Different types of AdS (gauge) fields}

\paragraph{Massive fields}

For $\Delta$ generic the gauge invariance is purely algebraic and can be completely removed by a proper gauge condition ${\bar S^\dagger}{}^{\alpha}\Phi=0$. Such fields are called massive and the full set of constraints/equations of motion describing irreducible field reads as:
\begin{gather}
\label{me1}
T^{ij} \Phi = 0, \qquad N_i{}^j \Phi = 0, \; i<j, \qquad N_i \Phi = s_i \Phi,\\
{\bar S^\dagger}{}^{i}\Phi=0,\\
(N_X + \Delta) \Phi = 0,\\
\label{me4}
\Box \Phi = 0, \qquad S^i \Phi = 0.
\end{gather}

\paragraph{(Partially) massless fields} 

If $\Delta,p,s_i$ are such that there exists $t \in \left\{ 1, 2, \ldots, s_p - s_{p+1} \right\}$ satisfying  $\Delta = t + p - s_p$ the gauge transformation is not completely algebraic and the field is a genuine gauge field called (partially) massless. In particular for $t = 1$ it is called massless.

Note that for $t = 1$ (i.e. massless field) and $s_1 = s_2 = \ldots = s_p$ the field is associated to a unitary $o(d-1,2)$-module~\cite{Metsaev:1995jp}. The important technical point is that thanks to the constraint algebra for a unitary massless field ``all tangent constraints'' hold:
\begin{equation}
  {\bar S^\dagger}{}^{i}\Phi=0.
\end{equation}

\paragraph{Critical fields}

Among the fields we consider there are so called \emph{critical}. They correspond to cases where $\Delta = \frac{d}{2} - \ell, \; \ell = 1, 2, 3, \ldots$. These are fields whose space of solutions contains a submodule of the form $(X^2)^\ell \Phi^+$. Such fields are of particular interest because they lead to a nontrivial conformal equations on their leading boundary values while noncritical fields correspond to off-shell boundary values.

In particular, all (partially) massless fields in odd-dimensional AdS space are critical. Among massive fields only those with $\Delta = \frac{d}{2} - \ell, \; \ell = 1, 2, 3, \ldots$ are critical.

Boundary values of unitary massless fields had been described at the level of equations of motion in \cite{Chekmenev:2015kzf}. In this work we expand the description to massive fields and propose a systematic construction of the respective Lagrangians.
In what follows we restrict ourselves to massive or unitary massless mixed-symmetry fields or generic totally symmetric fields.

\section{Conformal fields as boundary values} \label{sec:boundary_values}

A useful way to describe a conformal boundary of AdS space is to identify it with rays of the null-cone.  This can be equipped with the metric by identifying conformal boundary with a section of the null-cone and pulling back the ambient metric to the section.~\footnote{Identification of the conformally flat space as a space of null rays in the ambient space is extensively used in describing conformal fields since~\cite{Dirac:1936fq,Marnelius:1978fs}. See e.g.~\cite{Weinberg:2010fx,Costa:2011mg} for more recent developments}.
In what follows we chose to work with Minskowski metric on the boundary but the formalism can be naturally generalised~\cite{Grigoriev:2018mkp} to generic conformally flat boundary metrics. 

To study boundary values the strategy is to consider the ambient system in the vicinity of the section of a null-cone  and to identify values of the ambient field on the section as
its boundary value.  It turns out that constraints/EOMs on the ambient field give rise to constraints on boundary values.
These can also be seen as obstructions to lift an unconstrained boundary field to the field subject to the constraints of the previous section. This approach was put forward in~\cite{Bekaert:2012vt,Bekaert:2013zya,Chekmenev:2015kzf,Grigoriev:2018mkp}, where it was shown that in this way one indeed arrives at a concise formulation (at the level of equations of motion) of the (generalized) Fradkin-Tseytlin fields on the boundary.

A crucial technical tool employed in~\cite{Bekaert:2012vt,Bekaert:2013zya,Chekmenev:2015kzf,Grigoriev:2018mkp} is the parent formulation approach which allows to perform the ambient construction in the fiber of a suitable fiber bundle rather than in the space-time.  This is achieved by replacing $X^A$ coordinates with formal variables $Y^A$ and bringing the entire system to geometrical 1st order (in space time) form. After this one can safely consider the system to be defined in generic coordinates on either AdS or conformal boundary.  Such description gives both manifestly local and manifestly conformal description. Moreover, it originates from BV-BRST framework and hence properly takes into account gauge systems. Here we closely follow the exposition of~\cite{Chekmenev:2015kzf} to which we refer for further details.

To be more precise, the parent reformulation  of the system \eqref{me1}-\eqref{me4} has the following form: 
\begin{gather}
  \label{first_line}
  \nabla \Phi = 0,\qquad
  ((Y + V) \cdot \d_Y + \Delta) \Phi = 0,\qquad
  (Y + V) \cdot \d_{P_i} \Phi = 0,\\
  P_i{} \cdot \d_{P_i} \Phi = s_i \Phi\qquad
  \d_{P_i} \cdot \d_{P_j} \Phi = 0,\qquad
  P_i \cdot \d_{P_j} \Phi = 0,\\
  \label{third_line}
  \d_Y \cdot \d_{P_i} \Phi = 0, \qquad \Box_Y \Phi = 0.
\end{gather}
where $\Phi(x,Y,P)$ is now defined on the conformal space with local coordinates $x^\mu$, $V^A$ are components of a fixed section, which we take to be $V^+=1,V^-=0,V^a=0$, and $\nabla$ is given by
\begin{equation}
	\nabla = \derham + \frac12 \omega^{AB} J_{AB},
\end{equation}
where $\derham = dx^\mu \frac{\partial}{\partial x^\mu}$ is the de Rham differential, $J_{AB}$ denote $o(d,2)$ generators in twisted representation
\begin{equation}
	J_{AB} = (Y_A + V_A) \frac{\partial}{\partial Y^B} + P_{i A} \frac{\partial}{\partial P_i^B} - (A \leftrightarrow B)
\end{equation}
and $\omega_{\mu}^{AB}$ are coefficients of the flat $o(d,2)$-connection such that $\Omega_{\mu+}^a$ is invertible and is identified as the background frame field. The convenient choice is
\begin{equation}
	\omega_+^a=-\omega_a^-=dx^a\,, ~~~ \omega^+_-=\omega_-^a=\omega^+_b=\omega^a_b=0
\end{equation}
so that $\nabla$ takes the form
\begin{equation}
  \nabla_a = \hat\d_a - (Y^++1) \frac{\d}{\d Y^a} + Y_a \frac{\d}{\d Y^-}- \sum_i P_i^+ \frac{\d}{\d P_i^a},
\end{equation}
where $\hat \d^a = \d^a + \sum\limits_i P_i^a \frac{\d}{\d P_i^-}$.


The equations~\eqref{first_line} determine the dependence on $Y^a,Y^+$ and hence have a unique solutions for a given initial value $\phi(x,P,Y^-)$. In terms of $\phi$ the remaining equations read as:
\begin{gather} 
\label{unitary_varphi_box}
	\tilde\Box \phi + \frac{\d}{\d u} \left( d - 2 \big( \Delta + u \frac{\d}{\d u} \big) \right) \phi = 0,
	\\
 \label{unitary_varphi_div}
	(\d_{p_i} \cdot \d) \phi + \frac{\d}{\d w_i} \left( d + n_i - \Delta - 1 - 2 u \frac{\d}{\d u} \right) \phi + \sum\limits_{j \ne i} \frac{\d}{\d w_j} (p_j \cdot \d_{p_i}) \phi = 0,
\\
\label{unitary_spin}
\left( n_i + n_{w_i} - s_i \right) \phi = 0,
\\
	(p_i \cdot \d_{p_j}) \phi + w_i \frac{\d}{\d w_j} \phi = 0, \quad i < j,
	\label{unitary_varphi_Young}
	\\
\label{unitary_varphi_trace}
	(\d_{p_i} \cdot \d_{p_j}) \phi - 2 u \frac{\d}{\d w_i} \frac{\d}{\d w_j} \phi = 0,
\end{gather}
where $n_{w_i}=w_i \frac{\d}{\d w_i}$, $\tilde\Box = \hat \d^a \hat \d_a$, $\hat \d^a = \d^a + \sum\limits_i p_i{}^a \frac{\d}{\d w_i}$ and we used the following notations:
$p_i^a \equiv P^a_i, w_i \equiv P_i^-$, $u \equiv Y^-$.

Equation \eqref{unitary_varphi_box} determines $u$-dependence of $\phi$ and imposes on $\phi_0 = \phi|_{u=0}$ the equation ${\tilde\Box}^\ell \phi_0 = 0$. Recall that for critical fields $\ell=\frac{d}{2}-\Delta$, where in the (partially-) massless case $\Delta=t+p-s_p$. Note that equation \eqref{unitary_varphi_box} doesn't determine coefficient before $u^\ell$ in terms of $\phi_0$. That corresponds to a subleading solution, describing a conserved current. But we are more interested on the leading solution and the equations it satisfies.

At $u=0$ equations~\eqref{unitary_varphi_div}-\eqref{unitary_varphi_trace} uniquely determine $\phi_0$ for a given initial data $\phi_{00}(x,p_i)=\phi_{0}|_{w_i=0}$ satisfying
\begin{equation} \label{lorentz-irr}
 (n_i-s_i)\phi_{00}=0\,, \qquad (\d_{p_i} \cdot \d_{p_j}) \phi_{00}=0\,,  \qquad (p_i \cdot \d_{p_j})\phi_{00}=0\quad i<j\,,
\end{equation} 
which are precisely the conditions that $\phi_{00}$, as a function in $x^a$, takes values in the irreducible module with weights $s_1, \ldots, s_{n-1}$ of the Lorentz $o(d-1,1)$ subalgebra of $o(d,2)$. In other words, there is a map $\pi: \phi_0 \mapsto \phi_0|_{w_i=0}$ which sends solutions of \eqref{unitary_varphi_div}-\eqref{unitary_varphi_trace} to the space of unconstrained fields with values in irreducible Lorentz tensors. In appendix~\bref{app:w_lift} we show that this map is bijective, i.e. given $\phi_{00}(x,p)$ satisfying~\eqref{lorentz-irr} there exists a unique $\phi_0(x,p,w)$ satisfying~\eqref{unitary_varphi_div}-\eqref{unitary_varphi_trace}.

Given that $\pi$ is bijective the equations induced on $\phi_{00}$ can be written as
\begin{equation}
\label{mixed-comp}
\begin{gathered}
 ({\tilde\Box}^\ell\phi_0)|_{w_i=0}=0\,,\qquad  \phi_0|_{w_i=0}=\phi_{00}\,,\\
  (\d_{p_i} \cdot \d) \phi_0 + \frac{\d}{\d w_i} \left( d + s_i - \Delta - i - \sum\limits_{j \le i} n_{w_j} \right) \phi_0 + \sum\limits_{i<j} (p_j \cdot \d_{p_i}) \frac{\d}{\d w_j} \phi_0 = 0\,,
\end{gathered}
\end{equation} 
where the equations in the second line are interpreted as the constraints determining the $w_i$-dependence in a unique way (see \cite{Chekmenev:2015kzf} and Appendix~\bref{app:w_lift} for details). These equations are by construction conformally invariant though the invariance is not manifest in this form.

In the case of massless fields, i.e. $t=1$ and $\Delta = 1 + p - s_p$, the equations on $\phi_{00}$ encoded in \eqref{mixed-comp} are invariant under the following gauge transformations:
\begin{equation}
  \delta \phi_{00} = \left( \sum_{\alpha} (p_\alpha \cdot \hat \d) \lambda^\alpha \right)\Bigg |_{w_i=0},
\end{equation}
where $\lambda^\alpha(x,p,w)$ is  itself determined in terms of the gauge parameter $\lambda^\alpha_{00}$
via $\lambda^\alpha_{00}=\lambda^\alpha|_{w_i=0}$
and the following equations
\begin{equation} \label{gauge_w_lift}
(\d_{p_i} \cdot \d) \lambda^\alpha + \frac{\d}{\d w_i} \left( d + \tilde s_i - \tilde \Delta - i - \sum\limits_{j \le i} n_{w_j} \right) \lambda^\alpha + \sum\limits_{i<j} (p_j \cdot \d_{p_i}) \frac{\d}{\d w_j} \lambda^\alpha = 0,
\end{equation} 
where $\tilde \Delta = \Delta - 1$, $\tilde s_\alpha = s_\alpha - 1$.

In the case of totally symmetric (partially)-massless field in the bulk (i.e. $n=2,\Delta = t + 1 - s$), the gauge transformation takes the form
\begin{equation}
	\delta \phi_{00} = ( \Pi (p \cdot \hat \d)^t \lambda )|_{w = 0},
\end{equation}
where $\Pi$ denotes projection to the traceless component. In this case \eqref{gauge_w_lift} takes the following simple form
\begin{equation}
  (\d_p \cdot \d) \lambda + \frac{\d}{\d w} \left( d + s - \Delta - 1 - n_w \right) \lambda  = 0.
\end{equation}

\section{Conformal Lagrangians} \label{sec:lagrangians}

It turns out the equations on $\phi_{00}$ encoded in~\eqref{mixed-comp} has the same tensor structure as $\phi_{00}$ itself and hence have a chance to be Euler-Lagrange for some Lagrangians. As we are going to see in this section this is indeed the case. 

To see this let us consider the operator $\mathcal A = \pi \circ \tilde\Box^\ell \circ \pi^{-1}: \phi_{00} \mapsto (\tilde\Box^\ell \pi^{-1} \phi_{00})|_{w_i=0}$. In terms of $\mathcal A$ the first equation in \eqref{mixed-comp} take the following form $\mathcal A \phi_{00} = 0$. Next we introduce the formal inner product  
\begin{equation} \label{inner_product}
  \bfinner{ \phi}{\chi} = \int \dif^d x \langle \phi, \chi \rangle,
\end{equation}
where $\langle\cdot,\cdot\rangle$ is the standard inner product on polynomials in $p^a_i$ determined by the Minkowski metric $\eta_{ab}$. For instance, the corresponding formal conjugation rules read as:
\begin{equation} \label{conjugation}
x^\dag = x, \qquad \d_a{}^\dag = -\d_a, \qquad p_i^a{}^\dag = \eta^{ab} \frac{\d}{\d p_i^b}.
\end{equation}
Note that the inner product restricts to the subspace~\eqref{lorentz-irr} of Lorentz irreducible tensor fields.

We claim that $\mathcal A$ preserves the space~\eqref{lorentz-irr} of irreducible Lorentz tensors and is formally symmetric with respect to the above inner product.
This guaranties that equations $\mathcal A \phi_{00} = 0$ follow from the Lagrangian:
\begin{equation} \label{Lagrangian}
L = \langle \phi_{00}, \mathcal A \phi_{00} \rangle
= \langle \phi_{00}, (\tilde\Box^\ell \phi_{0})|_{w_i=0} \rangle
= \langle \phi_{0}, \tilde\Box^\ell \phi_0 \rangle|_{w_i=0},
\end{equation}
where $\phi_{00}$ is an irreducible Lorentz tensor, $\phi_0 = \pi^{-1} \phi_{00}$ is its unique lift via the last equation in \eqref{mixed-comp}. 

Moreover, the Lagrangian is gauge invariant. Indeed, equation on $\phi_{00}$ is gauge invariant: $\mathcal A \delta\phi_{00} = 0$ so that modulo total derivatives
\begin{equation}
\delta \langle \phi_{00}, \mathcal A \phi_{00} \rangle = \langle \delta \phi_{00}, \mathcal A \phi_{00} \rangle + \langle \phi_{00}, \mathcal A \delta \phi_{00} \rangle = 0
\end{equation}
because $\mathcal A$ is symmetric. 

In the rest of this section we demonstrate that $\mathcal A$ indeed preserves \eqref{lorentz-irr} and is formally symmetric there.

\subsection{Invariance of the Lorentz irreducible subspace}

First of all we demonstrate that the space of irreducible Lorentz tensors is invariant under $\mathcal A$. It other words, if $\phi_{00}$ satisfies \eqref{lorentz-irr} then $(\tilde\Box^\ell \phi_0)|_{w = 0}$ also does so provided $\phi_0$ is constructed as above.
For this it is sufficient to show that if $\phi_0$ is a solution of \eqref{unitary_varphi_div}-\eqref{unitary_varphi_trace} at $u=0$, then $\psi_0 = \tilde\Box^\ell \phi_0$ satisfies \eqref{unitary_spin}-\eqref{unitary_varphi_trace} at $u=0$. Let us write down explicitly the system of equations on $\phi_0$.
\begin{equation} \label{w_system_short}
  (d - \Delta - 1) \frac{\d}{\d w_i} \phi_0 + \divw^i \phi_0 =
  t^{i j} \phi_0 =
  (n_{p_i} + n_{w_i} - s_i) \phi_0 =
  \youngw_i{}^j \phi_0 = 0 \quad i < j,
\end{equation}
where
\begin{equation}
\begin{gathered}
  \divw^i \coloneqq (\d_{p_i} \cdot \hat \d) \equiv (\d_{p_i} \cdot \d) + \sum_j \frac{\d}{\d w_j} (p_j \cdot \d_{p_i}),\\
  \youngw_i{}^j \coloneqq (p_i \cdot \d_{p_j}) + w_i \frac{\d}{\d w_j},\qquad\quad
  t^{i j} \coloneqq (\d_{p_i} \cdot \d_{p_j}).
\end{gathered}
\end{equation}

We need to show that
\begin{equation}
	t^{i j} \psi_0 = (n_{p_i} + n_{w_i} - s_i) \psi_0 = \youngw_i{}^j \psi_0 = 0 \quad i < j.
\end{equation}
It follows from simple algebra that
\begin{gather}
  \comm*{n_{p_i} + n_{w_i} - s_i}{\tilde\Box} =
  \comm*{\youngw_i{}^j}{\tilde\Box} = 0,\\
  \comm*{t^{i j}}{\tilde\Box^\ell} = 2 \ell \tilde\Box^{\ell-1} \frac{\d}{\d w_i} \left( \divw^j + \big(\frac{d}{2} + \ell - 1\big) \frac{\d}{\d w_j} \right)  + (i \leftrightarrow j).
\end{gather}
The expression in parenthesis is exactly the operator in the first equation in \eqref{w_system_short} because for critical fields $\ell = \frac{d}{2} - \Delta$.

\subsection{Formal symmetry}

There remains to show that $\mathcal A = \mathcal A^\dag$. To this end let us observe first that elements of the form $(p_i \cdot p_j) \chi$ are orthogonal to $\phi_{00}$ because $(p_i \cdot p_j)^\dag = (\d_{p_i} \cdot \d_{p_j})$. So for any $\chi_{00}, \phi_{00}$ satisfying \eqref{lorentz-irr} one has
\begin{equation}
\langle \chi_{00}, (\tilde\Box^\ell \phi_0)|_{w_i=0} \rangle
= \Big\langle \chi_{00}, \Big( \Box + 2 \sum_i (p_i \cdot \d) \frac{\d}{\d w_i} \Big)^\ell \phi_{0}\bigg|_{w_i=0} \Big\rangle,
\end{equation}
where $\phi_0 = \pi^{-1} \phi_{00}$.

Then we recall the fact proved in \cite{Chekmenev:2015kzf} that equations \eqref{w_system_short} can be solved order by order in $\mathbb Z_{\ge 0}$-grading of weighted powers of $w_i:$ $\deg w_{n-1} = 1$, $\deg w_{n-2} = s_{n-1} + 1$, $\deg w_{n-3} = s_{n-2} \deg w_{n-2} + 1$ and so on: $\deg w_{i-1} = s_i \deg w_i + 1$ (see Appendix \bref{app:w_lift} for more details). Furthermore, the coefficient in front of $(w_1)^{k_1} \ldots (w_{n-1})^{k_{n-1}}$ in $\phi_0$ has the form $\mathcal O_{k_1 \ldots k_{n-1}} \phi_{00}$, where $\mathcal O_{k_1 \ldots k_{n-1}}$ is a linear differential operator of order $k_1+\ldots+k_{n-1}$ and has homogeneity $-k_i$ in $p_i$.

It follows that $\left.\left( \Box + 2 \sum_i (p_i \cdot \d) \frac{\d}{\d w_i} \right)^\ell \phi_{0}\right|_{w_i=0}$ has the form $\mathcal P \phi_{00}$, where $\mathcal P$ is a polynomial in $\Box, (p_k \cdot \d), (\d_{p_k} \cdot \d), (p_j \cdot \d_{p_i}) \; j>i$ such that it is a homogeneous differential operator of order $2\ell$ in $x^a$ that  preserves homogeneity in $p_k$. But any such operator is symmetric on the subspace of solutions of \eqref{lorentz-irr} with respect to the inner product $\langle \cdot , \cdot \rangle$ described above. Indeed, the linear span of $\Box, (p_k \cdot \d), (\d_{p_k} \cdot \d), (p_j \cdot \d_{p_i}) \; j>i$ is closed under commutator. Let us introduce ordering
\begin{equation}
(p_j \cdot \d_{p_i}) < \Box < (p_k \cdot \d) < (\d_{p_k} \cdot \d) < (p_{k+1} \cdot \d) < (\d_{p_{k+1}} \cdot \d) \quad j>i.
\end{equation}
After reordering $\mathcal P$ would become a sum of terms with $(p_i \cdot \d_{p_j})$ at the left and terms of the form $\Box^{k_0} (p_1 \cdot \d)^{k_1} (\d_{p_1} \cdot \d)^{k_1} \dots (p_{n-1} \cdot \d)^{k_{n-1}} (\d_{p_{n-1}} \cdot \d)^{k_{n-1}}$ with $k_0+k_1+\ldots+k_{n-1}=\ell$. The former ones produce terms orthogonal to $\phi_{00}$ because $(p_j \cdot \d_{p_i})^\dag = (p_i \cdot \d_{p_j}) \; i<j$ while the later ones are obviously symmetric under~\eqref{conjugation}.

\subsection{Examples}

\subsubsection{''Hook''-type field}

As an illustration of the construction let us consider the simplest mixed-symmetry field, the so-called ``hook''-type field which correspond to $d=4, s_1=2, s_2=1, p=1$. This field has a symmetry type described by Young diagram $(2,1)$, with the gauge parameter described by Young diagram $(1,1)$. In this case $\Delta = 1 + p - s_p = 0$, $\ell = \frac{d}{2} - \Delta = 2$.

Let us introduce notations for the coefficients of $\phi_0$ as follows:
\begin{equation}
\phi_0 = \phi_{00}
+ w_1 \phi_{10} + w_2 \phi_{01}
+ \frac12 (w_1)^2 \phi_{20} + w_1 w_2 \phi_{11}
+ \frac12 (w_1)^2 w_2 \phi_{21}.
\end{equation}

The subsystem \eqref{w_lift_subsystem} determining the $w_i$-dependence of $\phi_0$ read as
\begin{equation} \label{w_lift_2_1}
\begin{array}{ccc}
\text{degree} & w & \text{equation} \\
\begin{gathered} 0 \\ 1 \\ 2 \\ 3 \\ 4 \\ 5 \end{gathered} &
\begin{gathered} 1 \\ w_2 \\ w_1 \\ w_1 w_2 \\ (w_1)^2 \\ (w_1)^2 w_2 \end{gathered} &
\begin{aligned}
                 \phi_{00} \\
(\Delta + 2 - d) \phi_{01} &= (\d_{p_2} \cdot \d) \phi_{00} \\
(\Delta     - d) \phi_{10} &= (\d_{p_1} \cdot \d) \phi_{00} + (p_2 \cdot \d_{p_1}) \phi_{01} \\
(\Delta     - d) \phi_{11} &= (\d_{p_1} \cdot \d) \phi_{01} \\
(\Delta + 1 - d) \phi_{20} &= (\d_{p_1} \cdot \d) \phi_{10} + (p_2 \cdot \d_{p_1}) \phi_{11} \\
(\Delta + 1 - d) \phi_{21} &= (\d_{p_1} \cdot \d) \phi_{11} \\
\end{aligned}
\end{array}
\end{equation}
It follows from \eqref{unitary_varphi_Young} that $\phi_{21} = 0$ so that
\begin{multline}
( \tilde\Box^2 \phi_0 )|_{w=0} =
\Box^2 \phi_{00}
+ 4 \Box (p_1 \cdot \d) \phi_{10}
+ 4 \Box (p_2 \cdot \d) \phi_{01}
+\\+
4 (p_1 \cdot \d)^2 \phi_{20}
+ 8 (p_1 \cdot \d) (p_2 \cdot \d) \phi_{11}
+ \ldots,
\end{multline}
where ellipses denote terms that are in the images of $(p_j \cdot p_j)$ and $(p_j \cdot \d_{p_i}) \; i<j$. These terms ensure that 
the expression in the RHS is Lorentz irreducible.

Using \eqref{w_lift_2_1} we express $\phi_{ij}$ in terms of $\phi_{00}$ and substitute it in the last equation.
So according to \eqref{Lagrangian} the Lagrangian is
\begin{multline}
  L =
  \Big\langle
  \phi_{00},
  \Big(
  \Box^2
  - \Box (p_1 \cdot \d) (\d_{p_1} \cdot \d)
  - \frac52 \Box (p_2 \cdot \d) (\d_{p_2} \cdot \d)
  \\
  + \frac53 (p_1 \cdot \d) (\d_{p_1} \cdot \d) (p_2 \cdot \d) (\d_{p_2} \cdot \d)
  + \frac13 (p_1 \cdot \d)^2 (\d_{p_1} \cdot \d)^2
  \Big)
  \phi_{00}
  \Big\rangle
\end{multline}
In components ($\phi_{00} = p_1^a p_1^b p_2^c \phi_{abc}$, where $\phi_{abc} = \phi_{bac}$, $\phi_{abc} + \phi_{acb} + \phi_{bca} = 0$):
\begin{equation}
\frac12 L =
\phi^{abc} \Box^2 \phi_{abc}
+ 2 \d_{e} \phi^{ebc} \Box \d^f \phi_{fbc}
+ \frac52 \d_e \phi^{abe} \Box \d^f \phi_{abf}
+ \frac32 \d_a \d_b \phi^{abc} \d^e \d^f \phi_{efc},
\end{equation}
which reproduces the special case of the general Larangian proposed by Vasiliev~\cite{Vasiliev:2009ck}. The explicit form of this Lagrangian of  the ``hook'' field was obtained in  \cite{Alkalaev:2012ic} starting from the Lagrangian~\cite{Brink:2000ag} of the respective ``hook''-type field on AdS.



\subsubsection{Totally-symmetric fields}

In order to make connections to the literature let us consider the case of symmetric fields.

Equations on leading boundary value emerges for critical $\Delta = \frac{d}{2} - \ell, \; \ell \in \mathbb Z^{>0}$.
They are gauge invariant for even $d$ and $\Delta = 1, \dots, 2-s$, which corresponds to the case of (partially) massless fields. 
For other critical values of $\Delta$ and/or odd $d$ it describes massive fields. Thus critical values of $\Delta$ for even $d$ can be grouped as follows:
\begin{equation}
	\Delta
= \overbrace{\frac{d}{2} - 1, \ldots, 2}^{\text{massive}},
  \overbrace{\vphantom{\frac{d}{2}} 1, \ldots, 3-s}^{\text{partially massless}},
  \overbrace{\vphantom{\frac{d}{2}} 2-s}^{\text{massless}},
  \overbrace{\vphantom{\frac{d}{2}} 1-s, \ldots, -\infty}^{\text{massive}}.
\end{equation}
They are called \emph{special}, \emph{part-short}, \emph{short} and \emph{long} respectively in \cite{Metsaev:2016oic}.

The solution to the second equation of \eqref{mixed-comp} is
\begin{equation}
	\phi_0 = \sum_{k = 0}^s \varphi_k w^k,\qquad
  \varphi_0 = \phi_{00}, \quad (\frac{d}{2} - 1 + s - k + \ell) \varphi_k = - (\d_p \cdot \d) \varphi_{k - 1}.
\end{equation}

\begin{equation}
	\cA \phi_{00} = (\tilde \Box^\ell \phi_0)|_{w = 0} = \sum_{k = 0}^\ell \binom{\ell}{k} \Box^{\ell - k} 2^k (p \cdot \d)^k \varphi_k + \dots,
\end{equation}
where ellipses denote traceful terms that cancel in the Lagrangian.
Now \eqref{Lagrangian} takes the following form:
\begin{multline}
	L
= \sum_{k = 0}^{\min(s, \ell)} \binom{\ell}{k} 2^k \langle \phi_{00}, \Box^{\ell - k} (p \cdot \d)^k \varphi_k \rangle
=
\\
=
\sum_{k = 0}^{\min(s, \ell)} \binom{\ell}{k} (-2)^k \langle (\d_p \cdot \d)^k \phi_{00}, \Box^{\ell - k} \varphi_k \rangle.
\end{multline}
Up to an overall number this is exactly $\mathcal L$ from eq. (3.2) of \cite{Metsaev:2016oic}. It's gauge invariant for even $d$ and $1 \le \Delta \le 2 - s$.

\section{Ambient form of the Lagrangians}\label{sec:ambient-form}

Let us return to the parent description~\eqref{first_line}-\eqref{third_line} of the conformal fields.   Equations \eqref{first_line} are first order in $y^a, Y^+, P_i^+$ and have a unique solution for a given initial data $\phi(x|p,Y^-,P^-_i) \equiv \phi(x|p, u, w_i)$. The remaining equations in terms of $\phi$ are equations \eqref{unitary_varphi_box}-\eqref{unitary_varphi_trace}. If supplemented with a gauge equivalence~\cite{Bekaert:2013zya,Chekmenev:2015kzf} 
\begin{equation}
\label{gauge-equiv}
\Phi\sim \Phi +(Y+V)^{2\ell}\beta    \,,
\end{equation} 
where $\beta$ is subject to the analog of \eqref{first_line}-\eqref{third_line}, this system describes the on-shell conformal field. Indeed, the above gauge equivalence eliminates the subleading boundary value leaving us with the on-shell leading one.

It turns out that a slight modification of~\eqref{first_line}-\eqref{third_line} is an off-shell system which is useful in constructing the Lagrangian formulation. More precisely, the modified system reads as
\begin{gather}
  \label{conf_1}
  \nabla \Phi = 0,\qquad
  ((Y + V) \cdot \d_Y + \Delta) \Phi = 0,\qquad
  (Y + V) \cdot \d_{P_i} \Phi = 0,\\
  \label{conf_2}
  P_i{} \cdot \d_{P_i} \Phi = s_i \Phi\qquad
  \d_{P_i} \cdot \d_{P_j} \Phi = 0,\qquad
  P_i \cdot \d_{P_j} \Phi = 0,\\
  \label{conf_3}
  \d_Y \cdot \d_{P_i} \Phi = 0, \qquad \Box_Y \Phi = (Y + V)^{2 (\ell - 1)} \alpha,
\end{gather}
where an extra independent field $\alpha$ has been introduced.  $\alpha$ is an auxiliary field because  $(Y+V)^{2(\ell-1)}$ is an invertible element so that $\alpha$ is determined by $\Phi$. If we subject $\Phi$ to gauge equivalence~\eqref{gauge-equiv} and $\alpha$ to the induced equivalence relation
the resulting system describes off-shell conformal field $\phi_{00}(x,p)=\Phi|_{Y=0,P^{\pm}_i=0}$ subject to no equations (besides algebraic conditions~\eqref{lorentz-irr}).

In terms of the initial data $\phi(x|p_i,Y^-,P^-_i) \equiv \phi(x|p_i, u, w_i)$ equations \eqref{conf_2}-\eqref{conf_3} results in the system  \eqref{unitary_varphi_box}-\eqref{unitary_varphi_trace}, where  \eqref{unitary_varphi_box} is replaced with
\begin{equation} \label{box_with_alpha}
	\tilde\Box \phi + \frac{\d}{\d u} \left( d - 2 \big( \Delta + u \frac{\d}{\d u} \big) \right) \phi = (2 u)^{\ell - 1} \alpha|_{y^a = Y^+ = P^+_i = 0}.
\end{equation}

As we have seen, at $u=0$ equations~\eqref{unitary_varphi_div}-\eqref{unitary_varphi_trace} uniquely determine $\phi_0$ for a given initial data $\phi_{00}(x,p_i)=\phi_{0}|_{w_i=0}$ satisfying \eqref{lorentz-irr}. Then solving \eqref{box_with_alpha} order by order in $u$ we find that $C \tilde \Box^\ell \phi_0 = \alpha_{Y = P_i^+ = 0}$ for some nonvanishing constant $C$ and hence  that $\alpha|_{Y = P^\pm_i = 0} = C \mathcal A \phi_{00}$, where  $\mathcal A$ was defined in Section \bref{sec:lagrangians}. Finally, the Lagrangian~\eqref{Lagrangian} can be written (up to a coefficient) as
\begin{equation}
	\langle \Phi|_{Y = P^\pm_i = 0}, \alpha|_{Y = P^\pm_i = 0} \rangle = \langle \Phi|_{P^\pm_i = 0}, \alpha|_{P^\pm_i = 0} \rangle|_{Y = 0},
\end{equation}
where the inner product in the RHS is extended by linearity to formal series in $Y$.

It turns out that the expression for the Lagrangian can be rewritten in a manifestly $o(d,2)$-invariant form as
\begin{equation}
\label{manifest}
	\langle \Phi|_{P^\pm_i = 0}, \alpha|_{P^\pm_i = 0} \rangle|_{Y = 0} = \langle \Phi, \alpha \rangle|_{Y = 0} \equiv \left.\frac{\langle \Phi,  \Box_Y \Phi \rangle}{(Y+V)^{2(\ell-1)}}\right|_{Y=0},
\end{equation}
where by some abuse of notations the inner product is extended to the one determined by $\eta_{AB}$ on polynomials in all $P^A_i$. It is clear that for $P^{\pm}_i$-independent elements this inner product reduces to the initial one. 
To check~\eqref{manifest} observe that terms in $\alpha$ proportional to $P^-_i$ do not contribute because thanks to the last equation in \eqref{conf_1} one has
\begin{equation}
\inner{\Phi}{P_i^-\Psi}=\inner{\dl{P_i^+}\Phi}{\Psi}=-\inner{Y\cdot \d_{P_i}\Phi}{\Psi}
\end{equation}
that vanishes at $Y=0$. Similarly, terms in $\Phi$ proportional to $P_i^-$ also do not contribute because 
$\Box_Y (Y \cdot \d_{P_i}) \Phi = (Y \cdot \d_{P_i}) \Box_Y \Phi + 2 (\d_Y \cdot \d_{P_i}) \Phi$ and $\d_Y \cdot \d_{P_i} \Phi=0$ according to~\eqref{conf_3}. 
Furthermore, setting to zero terms in $\alpha,\Phi$ which are proportional to $P^-$ one concludes that terms proportional to $P^+$ do not contribute either because they may have nonvanishing inner product  with $P^+$-dependent elements only.

Finally, the Lagrangian \eqref{Lagrangian} can be written (up to a coefficient) as
\begin{equation}
\label{parent-ambient}
	L = \left.\frac{\langle \Phi,  \Box_Y \Phi \rangle}{(Y+V)^{2(\ell-1)}}\right|_{Y=0},
\end{equation}
where $\Phi$ is subject to the off-shell system \eqref{conf_1}-\eqref{conf_3}. This expression can be formally rewritten in the purely ambient terms as
\begin{equation}
\label{ambient-lagrangian}
	\tilde L = \frac{\langle \tilde\Phi,  \Box \tilde\Phi \rangle}{X^{2(\ell-1)}}\,.
\end{equation}
This description can be obtained from the above one by formally replacing $Y^A+V^A$ with $X^A$ and dropping the first equation in~\eqref{conf_1}.\footnote{More details on the relation between purely ambient description and its parent reformulation can be found in~\cite{Bekaert:2012vt}.} However, to give this purely ambient  Lagrangian a precise meaning one has to specify functional class in $X^A$ and interpret it as a Lagrangian in $d$ rather than $d+2$ dimensions. The Lagrangian description based on ~\eqref{parent-ambient} can be seen as a constructive way to give a precise meaning to the purely ambient  Lagrangian~\eqref{ambient-lagrangian}

\section{Conclusions}

In this work we have proposed a simple generating procedure for Lagrangians of a wide class of mixed-symmetry type conformal fields.  The class involves totally symmetric fields, conformal fields associated to unitary mixed-symmetry fields in AdS as well as generic massive fields.

It seems that the construction is also applicable to conformal fields associated to nonunitary mixed-symmetry fields on AdS but proving this requires extra technical steps which we leave for a future work. Mention also that the proposed Larangians have something in common with ordinary derivative Lagrangians proposed in~\cite{Metsaev:2007fq,Metsaev:2007rw} for totally symmetric fields. 

An important and more conceptual point is to understand what the formal symmetry of the kinetic operator means in terms of the bulk dynamics. This may also shed some light on the long-standing problem of constructing a proper Lagrangian description of generic mixed-symmetry fields on AdS. Note that Lagrangians for some special classes  of mixed-symmetry fields are available in the literature~\cite{Brink:2000ag,Zinoviev:2003ix,Alkalaev:2005kw}.

\section*{Acknowledgments}
We are grateful to R.~Metsaev for useful discussions. The work of
A.C. was supported by the Russian Science Foundation grant 18-72-10123 in association with the Lebedev Physical Institute. The work of M.G. was supported by the  RFBR grant 18-02-01024.

\appendix

\section{Existence of the lift} \label{app:w_lift}

We want to show that the system \eqref{w_system_short} is consistent and has a unique solution $\phi_0$ such that $\phi_0|_{w_i = 0} = \phi_{00}$ for any $\phi_{00}$ satisfying \eqref{lorentz-irr}.

From spin constraints it follows that $\phi_0$ is polynomial in $w$ with $w_i$ degree no more than $s_i$. So essentially it's the finite dimensional linear algebra problem about nonhomogeneous system of linear equations. We follow steps similar to Gaussian elimination: transform the system into row echelon form, indicate a subsystem providing a unique solution and then show that other equations are consistent with that solution.

\paragraph{Row echelon form}

Taking into account $w$-extended Young and spin constraints we get the equivalent system
\begin{equation} \label{w_lift_full_system}
	\divwfullasym^i \phi_0 = t^{i j} \phi_0 = (n_{p_i} + n_{w_i} - s_i) \phi_0 = \youngw_i{}^j \phi_0 = 0 \quad i < j,
\end{equation}
where
\begin{gather}
  \div^i \coloneqq (\d_{p_i} \cdot \d), \qquad
  \divwcenter^i \coloneqq \frac{\d}{\d w_i} \Big( d + s_i - \Delta - i - \sum\limits_{j \le i} n_{w_j} \Big), \qquad
  y_j{}^i \coloneqq {p_j \cdot \d_{p_i}},\\
	\divwfullasym^i \coloneqq \div^i + \divwcenter^i + \sum_{j > i} y_j{}^i \frac{\d}{\d w_j}.
\end{gather}

\paragraph{Subsystem}

Let us denote by $\phi_0|_{m_1 \dots m_{n-1}}$ coefficient before $w_1^{m_1} \dots w_{n-1}^{m_{n-1}}$.
Equation $(\divwfullasym^i \phi_0)|_{m_1 \dots m_{n-1}} = 0$ in components reads
\begin{equation} \label{div_components}
	\div^i \phi_{m_1\dots m_{n-1}} + \Big(d + s_i - \Delta - i - \sum_{k \le i} m_k - 1\Big) \phi_{\dots m_i+1\dots} + \sum_{k > i} y_k{}^i \phi_{\dots m_k+1\dots} = 0.
\end{equation}

Consider the subsystem
\begin{equation} \label{w_lift_subsystem}
	(\divwfullasym^i \phi_0)|_{0\,\dots\,0\,m_i\,\dots\,m_{n-1}} = 0 \qquad i = 1, \dots, n-1, \quad m_j = 0, \dots, s_j - \delta_j^i
\end{equation}
(equations with $m_i = s_i$ are trivial due to spin constrain). In this case
\begin{equation}
	d + s_i - \Delta - i - \sum_{k \le i} m_k - 1 = (\frac{d}{2} - i) + (s_i - 1 - m_i) + \ell > 0
\end{equation}
so \eqref{w_lift_subsystem} can be used to solve order by order in $w_i$ in the order defined by $\mathbb Z_{\ge 0}$-grading of weighted powers of $w_i:$ $\deg w_{n-1} = 1, \deg w_{i-1} = s_i 
\deg w_i + 1$. 

More concretely, in this way one first solves the equation \eqref{w_lift_subsystem} with $i = n-1$ in the subspace of $w_j$-independent elements with $j < i$. Then one uses the solution as the initial data for the equation with $i = n-2$ and solves it in the subspace of $w_j$-independent elements with $j < i$. And so on: solution of \eqref{w_lift_subsystem} with $i > j$ in the subspace of $w_k$-independent elements with $k \le i$ is used as the initial data for \eqref{w_lift_subsystem} with $i = j$ to get the solution in the subspace of $w_k$-independent elements with $k < i$.

\paragraph{Consistency}

By construction $\phi_0|_{m_1 \dots m_{n-1}} = \cP \phi_{00}$ where $\cP$ is a polynomial in $\div^i, y_j{}^i \; i < j$ such that $\deg_{p_i} \cP = - m_i$. So spin constraints are satisfied. Trace constraints follow from the algebra:
\begin{equation}
	\comm{\div^k}{t^{i j}} = 0, \qquad \comm{t^{ij}}{y_n{}^m} = \delta^i_n t^{mj} + (i \leftrightarrow j).
\end{equation}

For divergency-like constraint we use double induction by Young row number $i = n-1, \dots, 1$ and cell number $m_i = 0, \dots, s_i$.
Assume that equations
\begin{equation}
	(\divwfullasym^j \phi_0)|_{0 \dots 0\,m_i\,m_{i+1} \dots m_{n-1}} = 0 \quad j > i
\end{equation}
hold for some $i$ and $m_i$ and for arbitrary $m_{i+1}, \dots, m_{n-1}$.
Then acting with $\div^j$, $j > i$ on $(\divwfullasym^i \phi_0)|_{0 \dots 0 \, m_i \dots m_{n-1}} = 0$ (which is given), rearranging terms and using
\begin{equation}
\begin{aligned}
  &(\divwfullasym^i \phi_0)|_{0 \dots 0 \, m_i \dots m_k+1 \dots} = 0 \quad k \ge j,\\
  &(\divwfullasym^j \phi_0)|_{0 \dots 0 \, m_i \dots m_{n-1}} = 0,\\
  &(\divwfullasym^j \phi_0)|_{0 \dots 0 \, m_i \dots m_k+1 \dots} = 0 \quad k > i.
\end{aligned}
\end{equation}
we arrive at
\begin{equation}
	( d + s_i - \Delta - i - m_i - 1 ) (\divwfullasym^j \phi_0)|_{0 \dots 0 \, m_i+1 \, m_{i+1} \dots m_{n-1}} = 0.
\end{equation}

Analogously for $w$-extended Young constraints. Assume that equations
\begin{equation}
  (\youngw_j{}^k \phi_0)|_{0 \dots 0\,m_i\,m_{i+1} \dots m_{n-1}}.= 0 \quad i \le j < k
\end{equation}
hold for some $i$ and $m_i$ and for arbitrary $m_{i+1}, \dots, m_{n-1}$.
Acting with $y_j{}^k$, $k > j$ on $(\divwfullasym^j \phi_0)|_{0 \dots 0\,m_i \dots m_{n-1}} = 0$ and using
\begin{equation}
\begin{aligned}
  &(\divwfullasym^j \phi_0)|_{0 \dots 0 \dots m_j-1 \dots m_k+1 \dots} = 0,\\
  &(\divwfullasym^k \phi_0)|_{0 \dots 0\,m_i \dots m_{n-1}} = 0,\\
  &(\youngw_l{}^k \phi_0)|_{0 \dots 0 \, m_i \dots m_l+1 \dots} = 0 \quad j < l < k,\\
  &(\youngw_j{}^k \phi_0)|_{0 \dots 0 \, m_i \dots m_j \dots} = 0,\\
  &(\youngw_j{}^k \phi_0)|_{0 \dots 0 \, m_i \dots m_l+1 \dots} = 0 \quad l > j.
\end{aligned}
\end{equation}
we arrive at
\begin{equation}
  ( d + s_i - \Delta - i - m_i - 1 ) (Y_j{}^k \phi_0)|_{0 \dots 0 \, m_i+1 \, m_{i+1} \dots m_{n-1}} = 0.
\end{equation}
The base case $(\youngw_{n-2}{}^{n-1} \phi_0)|_{0 \dots 0 \, m_{n-1}} = y_{n-2}{}^{n-1} \phi_0|_{0 \dots 0 \, m_{n-1}}$ is true because $\phi_0|_{0 \dots 0 \, m_{n-1}} \propto (\div^{n-1})^{m_{n-1}} \phi_{00}$.
This ends the proof of consistency.

Note that from $\youngw_i{}^j \phi_0 = 0$ it follows that elements $\phi_0|_{\dots s_i \dots m_j \dots}$ with maximum $m_i = s_i$ and $m_j \ne 0$, $i < j$ vanish. Also, if $\phi_0|_{\dots m_i \dots m_j \dots} = 0$ then $\phi_0|_{\dots m_i-1 \dots m_j+1 \dots} = 0$. This implies that elements with $\sum_{k \ge i} m_k > s_i$ vanish.




\providecommand{\href}[2]{#2}\begingroup\raggedright\endgroup

\end{document}